\documentclass{aa}
\usepackage{txfonts}
\usepackage{graphicx}
\begin{document}

\title{Str\"{o}mgren metallicities for intermediate-age and old star clusters}

\author{Andr\'es E. Piatti\inst{1,2}\thanks{\email{andres.piatti@unc.edu.ar}}}

\institute{Instituto Interdisciplinario de Ciencias B\'asicas (ICB), CONICET-UNCUYO, Padre J. Contreras 1300, M5502JMA, Mendoza, Argentina;
\and Consejo Nacional de Investigaciones Cient\'{\i}ficas y T\'ecnicas (CONICET), Godoy Cruz 2290, C1425FQB,  Buenos Aires, Argentina\\
}

\date{Received / Accepted}

\abstract{We report results that show that the straightforwardly star cluster metallicities obtained from
Str\"omgren $vby$ photometry is age-dependent and need to be corrected for further use. This 
outcome arises from the comparison of [Fe/H] values derived from Str\"omgren photometry with those 
metallicities published in the literature for 26 Large and Small Magellanic Cloud star clusters, whose
ages range from $\sim$ 1 Gyr up to the known oldest globular clusters' ages in these galaxies. While
deriving mean star cluster metallicities we carried out a thorough selection of red giant branch 
candidates to comply with the Str\"omgren metallicity calibration validity regime. We paid attention to 
the effect of contamination by field stars, particularly of those that lie inside the star clusters' radii, 
distributed along the star cluster red giant branches, and with [Fe/H] values covering a similar range as
that for the selected stars. We found that the measured Str\"omgren metallicities are systematically 
more metal-poor than the published ones and that a quadratically age-varying function reproduces the
relative metallicity values with an overall uncertainty of $\sim$ 0.05 dex.  We finally performed a similar 
comparison relying on a  fully independent approach, that consisted in using theoretical red giant 
branches  of old globular clusters spanning [Fe/H] values from -2.0 up to 0.0 dex as standard ones. 
We then superimposed on to them the red giant branches of star clusters with ages in the range 
1.0 - 12.5 Gyr and estimated by interpolation their associated metallicities. The derived theoretical
relative metallicities follow a similar trend as a function of the star clusters' ages than that found from
observations of star clusters.
}
 

\titlerunning{}

\authorrunning{Andr\' es E. Piatti }

\maketitle

\markboth{Andr\' es E. Piatti: }{Str\"omgren metallicities of star clusters}

\section{Introduction}

Star clusters' metallicities have long been used as key parameters for unveiling the chemical evolution
of galaxies. Their metal abundances tell us about the efficiency of the gas mixing within a galaxy, the 
metal enrichment along the galaxy lifetime,  the infall of gas from galaxy interactions, etc. Alongside with
ages and positions (age-metallicity relationships, metallicity gradients), star clusters' metallicities have
helped us with understanding whether galaxies formed through outside-in or inside-out formation 
scenarios,  detecting the existence of bursting formation episodes that produced a sudden chemical 
enrichment, estimating the effectiveness of scattering star clusters, among others.

In this context, the red giant branch of old star clusters has been employed to estimate metallicities
from a variety of photometric systems, namely: Johnson $BV$ \citep{h68}, Johnson-Cousins $VI$ 
\citep{da90}, Washington $CT_1$ \citep{gs99}, near-IR $J,Ks$ bands \citep{frogeletal1983}, etc.
This is because the colors of the red giant branch stars show sensitivity to changes in metallicities. The 
result  is that the red giant branches appear shifted towards bluer regions in the color-magnitude diagram 
(CMD) as the star clusters are more metal-poor. In practice, a fixed magnitude level is adopted and the 
color differences at that magnitude level is obtained and transformed into [Fe/H] values. Nevertheless, 
having a long and well-populated red giant branch is advantageous, since it can be superimposed on to
the standard iso-abundance red giant branches and thus to obtain by interpolation a metallicity estimate.

These  iso-abundance red giant branches have usually been drawn from CMDs of old globular 
clusters, so that when using them for estimating metallicities of younger star clusters, the derived
[Fe/H] values result much more metal-poor because of the age-metallicity degeneracy. This
phenomenon has been studied from numerical simulation for the Johnson $VI$ photometric system by 
\citet{os15}, who showed that the metallicity derived from the star cluster's red giant branch stars can be 
in error by up to $\sim$ 0.5 dex, if the star cluster is younger than $\sim$ 6 Gyr old. \citet{getal03} also
derived an age-dependent metallicity correction for the Washington $CT_1$ photometric system, which
shows that [Fe/H] values of star clusters of $\sim$ 1 Gyr old need to be corrected by $\sim$ 0.7 dex 
in order to compute reliable metallicities.

The Str\"omgren $vby$ medium-bandwidth filters \citep{cm1976} have proved to be able to derive 
straightforwardly accurate metallicities values ([Fe/H]) for many stars in a star cluster field, provided the
photometric data are precise \citep[see, e.g.,][]{franketal2015,massarietal2016,gruytersetal2017}.
This is because of the index $m_1$ = $(v - b) - (b - y)$ has resulted to be a metallicity-sensitive one,
as judged by the available calibrations of it as a photometric proxy of iron abundances.  Particularly, the
semi-empirical calibration obtained by \citet{calamidaetal2007} has turned out to be the most robust one
for estimating metallicities of red giants \citep{adenetal2009, calamidaetal2009,arnaetal2010}. 
It was derived using high-dispersion spectroscopic data of red giants of Milky Way globular clusters 
and and alpha-enhanced isochrones transformed to the observational plane by using semi-empirical 
color-temperature relations.

In this work, we present Str\"omgren $vby$ photometry of intermediate-age Large and Small
Magellanic Cloud (LMC/SMC) star clusters aiming at investigating at what extend the Str\"omgren 
metallicities are affected by the age-metallicity degeneracy. If such a age-metallicity dependence
exists, it should arise from the comparison of the metallicities measured here from the Str\"omgren
indices with those accurate values available in the literature. The paper is organized as follows: 
Section 2 introduces the obtained Str\"omgren $vby$ images and describes their processing until 
obtaining the standardized photometric data sets. Sections 3 and 4 deal with the metallicity estimates 
and the aforementioned age-metallicity correction. Finally, Section 5 summarizes the main results of 
this work. We note that a metallicity calibration based on young stellar evolutionary 
models would be of great value. Such a calibration would consider the different evolutionary status of young stars ($\sim$ 1 Gyr) compared to old stars ($>$ 10 Gyr) in different 
Str\"omgren colors' planes.

\section{Str\"omgren photometric data}

We made use of Str\"omgren $vby$ images collected during the observing program SO2008B-0917
(PI: Pietrzy\'nski), conducted with the SOAR Optical Imager (SOI) attached to the 4.1m Southern
Astrophysical Research (SOAR) telescope (FOV = 5.25$\arcmin$$\times$5.25$\arcmin$, scale=
0.154$\arcsec$/px in binned mode). The observing program was executed during two different
epochs (17-19 December  2008 and 16-18 January 2009) under excellent image quality conditions
(typical FWHM $\sim$ 0.6$\arcsec$). We downloaded the observational material from the National 
Optical Astronomy Observatory (NOAO)  Science Data Management (SDM) 
Archives.\footnote{http //www.noao.edu/sdm/archives.php.}.  Table~\ref{tab:tab1} lists the log of 
observations for the studied LMC/SMC star clusters. 

\begin{table*}
\caption{Observing log and properties of the LMC/SMC star cluster sample.}
\label{tab:tab1}
\begin{tabular}{@{}lcccccccccccc}\hline\hline
Star cluster$^a$ & Date & \multicolumn{3}{c}{Exp. time (sec)} & \multicolumn{3}{c}{Airmass} & $E(B-V)$ & Age & Ref. & [Fe/H]$_{\rm adopted}$   & [Fe/H] \\
        &      &   $v$   &  $b$  &  $y$        &  $v$  & $b$  &  $y$ & (mag) & (Gyr) &  & (dex) &  (dex) \\\hline
        \multicolumn{13}{c}{LMC}  \\\hline
NGC\,1651  & 2008-12-18 & 500 & 200 & 120  & 1.54 & 1.53 & 1.53  & 0.07 & 2.00$\pm$0.20 & 1,2 & -0.70$\pm$0.10 &   -1.05$\pm$0.15 \\
NGC\,1795   & 2008-12-19 & 350 & 140 & 90    & 1.38 & 1.38 & 1.37  & 0.07 & 1.50$\pm$0.20 & 4,5 & -0.40$\pm$0.10 &  -0.90$\pm$0.15\\
NGC\,1846  & 2008-12-18 & 500 & 200 & 120  & 1.61 & 1.60 & 1.59  & 0.07 & 1.40$\pm$0.20 & 2,6 & -0.50$\pm$0.10 &  -0.90$\pm$0.15\\
NGC\,2155  & 2008-12-18 & 500 & 200 & 120  & 1.37 & 1.36 & 1.36 & 0.04 & 3.00$\pm$0.30 & 1,3 & -0.70$\pm$0.10 &  -1.00$\pm$0.10\\
SL\,8       & 2009-01-16      & 350 & 180 & 100  & 1.33 & 1.34 & 1.34 & 0.07 & 1.80$\pm$0.30 & 7 & -0.40$\pm$0.20 &   -0.85$\pm$0.10\\
SL\,363  & 2008-12-17       & 300 & 100 &  60   & 1.46 & 1.45 & 1.44 & 0.07 & 2.24$\pm$0.10 & 2 & -0.49$\pm$0.12 &  -0.90$\pm$0.15\\
SL\,388  & 2009-01-16       & 350 & 140 &  90   & 2.04 & 2.02 & 2.01 & 0.04 & 2.20$\pm$0.30 & 7 & -0.65$\pm$0.20 &  -0.95$\pm$0.10\\
SL\,509  & 2008-12-19       & 350 & 140 &  90   & 1.46 & 1.46 & 1.45 & 0.05 & 1.20$\pm$0.30 & 7 & -0.54$\pm$0.09 &  -1.10$\pm$0.15 \\
SL\,549   & 2009-01-18   & 350 & 140 &  90   & 1.84 & 1.83 & 1.82 & 0.05 & 1.70$\pm$0.30 & 8 & -0.70$\pm$0.20 &  -1.10$\pm$0.10 \\
SL\,555   & 2009-01-18      & 350 & 140 &  90   & 1.79 & 1.81 & 1.82 & 0.07 & 1.70$\pm$0.20 & 9 & -0.70$\pm$0.20 &   -1.10$\pm$0.10 \\
SL\,817  & 2009-01-17       & 450 & 180 & 110  &  2.12 & 2.10 & 2.08 & 0.07 & 1.50$\pm$0.30 & 7 & -0.41$\pm$0.05 &  -0.95$\pm$0.10\\
SL\,842  & 2009-01-18       & 400 & 180 & 100  &  2.01 & 2.06 & 2.09 & 0.05 & 2.20$\pm$0.30 & 7 & -0.60$\pm$0.20 &  -1.00$\pm$0.10\\
SL\,862   & 2009-01-18      & 400 & 180 & 100  &  2.04 & 2.02 & 2.02 & 0.07 & 1.80$\pm$0.30 & 7 & -0.47$\pm$0.07 &  -0.90$\pm$0.15  \\\hline
        \multicolumn{13}{c}{SMC} \\\hline
L\,6    & 2008-12-17 &500 & 300 & 180   & 1.91 & 1.89 & 1.88  &  0.03 & 3.30$\pm$0.70 & 10 & -1.07$\pm$0.17 &  -1.25$\pm$0.20\\
L\,19   & 2008-12-18 &500 & 300 & 180   & 1.51 & 1.50 & 1.50  & 0.03 & 2.10$\pm$0.20 & 10 & -0.75$\pm$0.10 &  -1.05$\pm$0.15\\
L\,27   & 2008-12-18 &500 & 300 & 180   & 1.46 & 1.45 & 1.44 & 0.03 & 2.10$\pm$0.20 & 10 & -1.14$\pm$0.06 & -1.50$\pm$0.15 \\\hline 
\end{tabular}

$^a$ Star cluster identifications are from \citet[][SL]{sl1963} and \citet[][L]{l1958}.

Ref. : (1) \citet{ketal07}; (2) \citet{goudfrooijetal14}; (3) \citet{pb16b}; (4) \citet{cetal14}; (5) \citet{getal10}; (6) \citet{p11a}; (7) 
\citet{betal98}; (8)  \citet{petal03b}; (9) \citet{getal03}; (10) \citet{petal05b}.

\end{table*}

We processed the images following the SOI's reduction prescriptions available at 
http://www.ctio.noao.edu/soar/content/soar-optical-imager-soi, so that we also used nightly zero and 
flat-field images. As for the photometry standardization, we employed observations of the following
Str\"omgren standard stars:  HD64, HD3417, HD12756, HD22610, HD57568, HD58489, HD66020,
TYC 7547-711-1, TYC 7548-698-1, TYC 7583-1011-1, TYC 7583-1622-1, TYC 7626-763-1,
TYC 8033-906-1, TYC 8067-207-1, TYC 8104-856-1 and TYC 8104-969-1 \citep{hm1998,p2005}.
They were observed in all the $vby$ filters at airmass spanning the range $\sim$ 1.02 - 2.20.
Standard stars were observed at a fixed airmass twice to allow them to be placed in the two different
CCDs arrayed by SOI, and thus to monitor their  individual responses. \citet{pb2019}
showed that the transformation coefficients obtained from magnitudes of standard stars placed
in each CCD are indistinguishable, so that we enlarged the sample of standard star magnitude
measurements by considering all of them irrespective of their positions in SOI. To convert instrumental 
magnitudes into standard ones, we  first  gathered all the information of the standard stars, and then 
fitted the following expressions:

\begin{equation}
v = v_1 + V_{\rm std} + v_2\times X_v + v_3\times (b-y)_{\rm std} + v_4\times m_{\rm 1 std},
\end{equation}
\begin{equation}
b = b_1 + V_{\rm std} + b_2\times X_b + b_3\times (b-y)_{\rm std}
\end{equation}
\begin{equation}
y = y_1 + V_{\rm std}  + y_2\times X_y + y_3\times (b-y)_{\rm std},
\end{equation}

\noindent where  $v_i$, $b_i$ and $y_i$ are the i-th fitted coefficients, and $X$ represents the effective
airmass. The resulting mean transformation coefficients are listed in Table 2 of \citet{piattietal2019c},
who used images from the same observing program to estimate metallicities of young star clusters
in the LMC/SMC from their supergiant populations.

We relied on the routine packages {\sc daophot}, {\sc allstar}, {\sc daomatch} and {\sc daomaster} 
\citep[stand-alone version,][]{setal90} to obtain point-spread-function (PSF) photometric data sets of 
the star clusters' fields and their associated uncertainties. In order to build the PSF of an image, we first
interactively selected nearly one hundred well-isolated, relatively bright, not-saturated stars, distributed
over the whole image area. A subsample of the nearly best forty PSF stars was used to build a preliminary 
PSF, that was applied to the image aiming at cleaning the entire PSF star sample 
from fainter neighbors. With the cleaned PSF stars, we constructed the final quadratically 
spatially-varying PSF for that image and computed aperture corrections that resulted in the range
-0.04 - -0.07 mag. The PSF was applied to the entire list of identified stellar sources in the image. The
resulting subtracted image was employed to identify new ones, which were fitted by the PSF 
simultaneously with those in the previous list. We took advantage of this procedure of enlarging the 
photometrically measured star sample by iterating it three times. Their standard magnitudes were 
obtained by inverting eqs. (1)-(3) and by entering with the computed instrumental magnitudes. The 
above described procedure was repeated for each of the images listed in Table~\ref{tab:tab1}. In the
subsequent analysis, we only kept those stellar sources with $\chi$ $<$ 2 and {\sc $|$sharpness$|$}
$<$ 0.5. $\chi$ is a {\sc daophot} robust estimate of the ratio: the observed pixel-to-pixel scatter 
from the model image profile divided by the expected pixel-to-pixel scatter from the image profile 
\citep[see Figure 28 in][]{sh1988}, while {\sc sharpness} is another image quality diagnostic defined as
the ratio of the height of the bivariate delta-function which best fits the brightness peak in the
original image to the height of the bivariate Gaussian function which best fits the peak. We adopted
the frequently used values of $\chi$ and {\sc sharpness} to exclude bad pixels, cosmic rays, galaxies 
and unrecognized double stars.
 
In order to quantify the uncertainties associated to the obtained photometry we followed the recipe
applied in previous studies of other subsets of star clusters observed during the same observing
program  \citep[see][]{pk2018,p18b}. The method consists in adding to an image synthetic stars with 
magnitudes and positions distributed similarly to those of the measured stars, and carrying out the 
photometry for the new image as described above. The synthetic stars represent nearly
5$\%$ of the measured stars, so that the original image stellar density is not significantly altered.  The 
resulting magnitudes for the synthetic stars are then compared with those used to create such stars.
The difference between them turned out to be typically equal to zero and in all the cases smaller than 
0.003 mag. We then decided to use as photometric errors the rms errors obtained from the comparison
of input and output synthetic star magnitudes.

\begin{figure}
\includegraphics[width=\columnwidth]{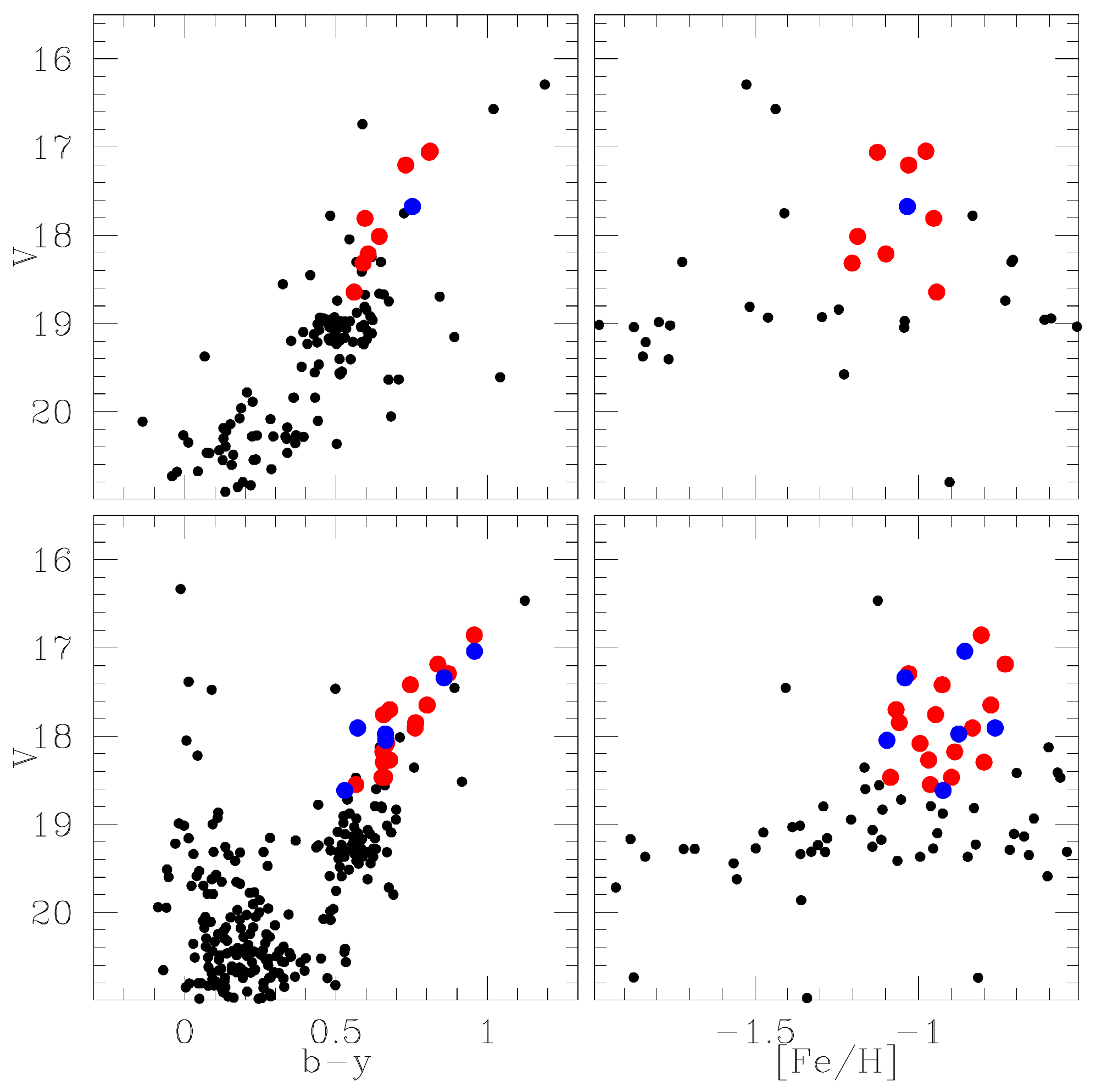}
\caption{$V$ vs. $b-y$ CMD for measured stars distributed inside the star cluster's radius (left panel)
and their metallicity ([Fe/H]) distribution as a function of the $V$ magnitude (right panel). Red and blue
large filled circles represent stars used to compute the star cluster's mean [Fe/H] value and those from
a reference star field with [Fe/H] values similar to those of the star cluster's stars, respectively (see text for
details). Top and bottom panels refer to NGC\,1651 and 1795, respectively.}
\label{fig:fig1}
\end{figure}

\begin{figure}
\includegraphics[width=\columnwidth]{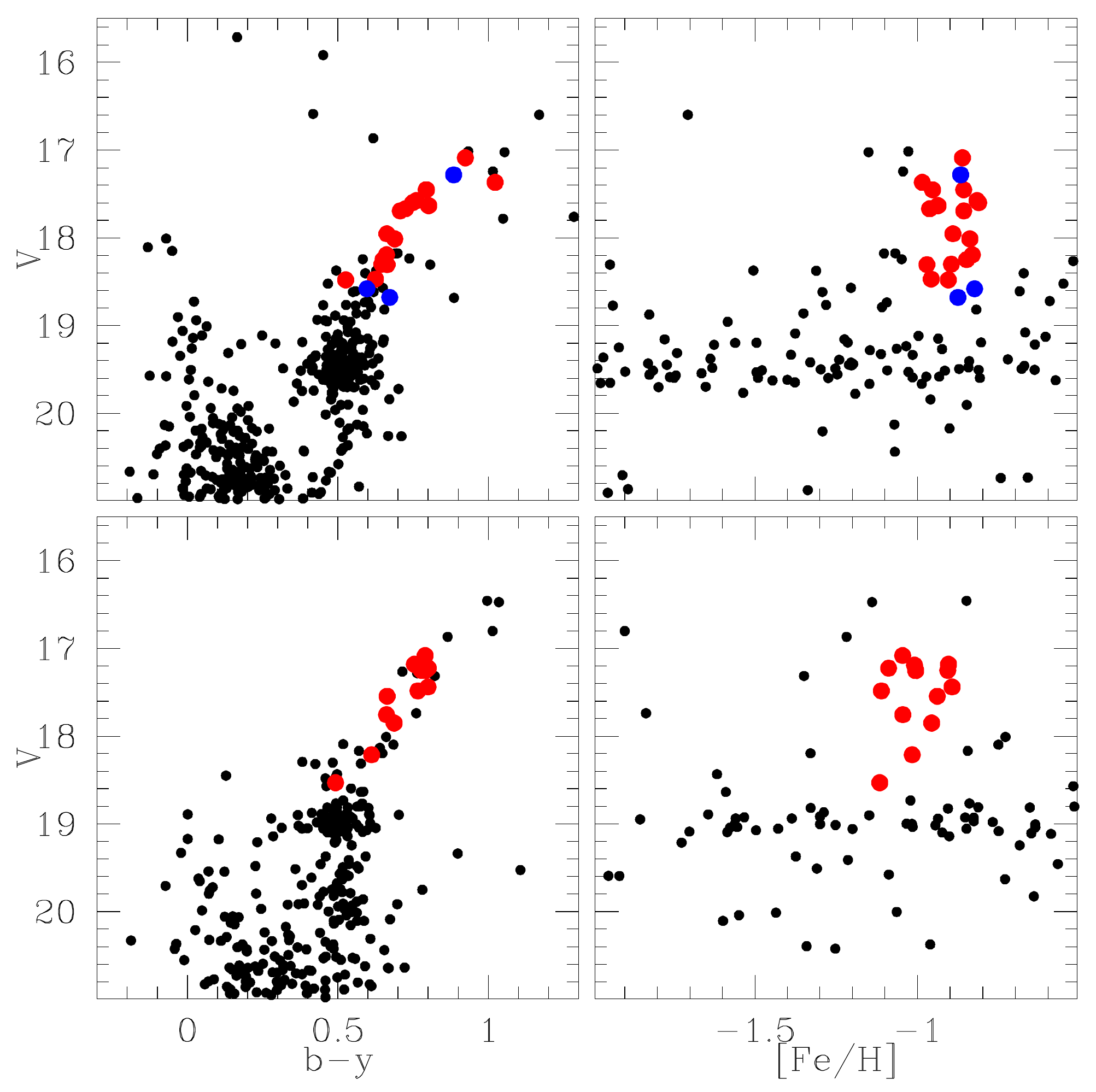}
\caption{Same as Figure~\ref{fig:fig1}, for NGC\,1846 (top panels) and 2155 (bottom panel).}
\label{fig:fig2}
\end{figure}

\begin{figure}
\includegraphics[width=\columnwidth]{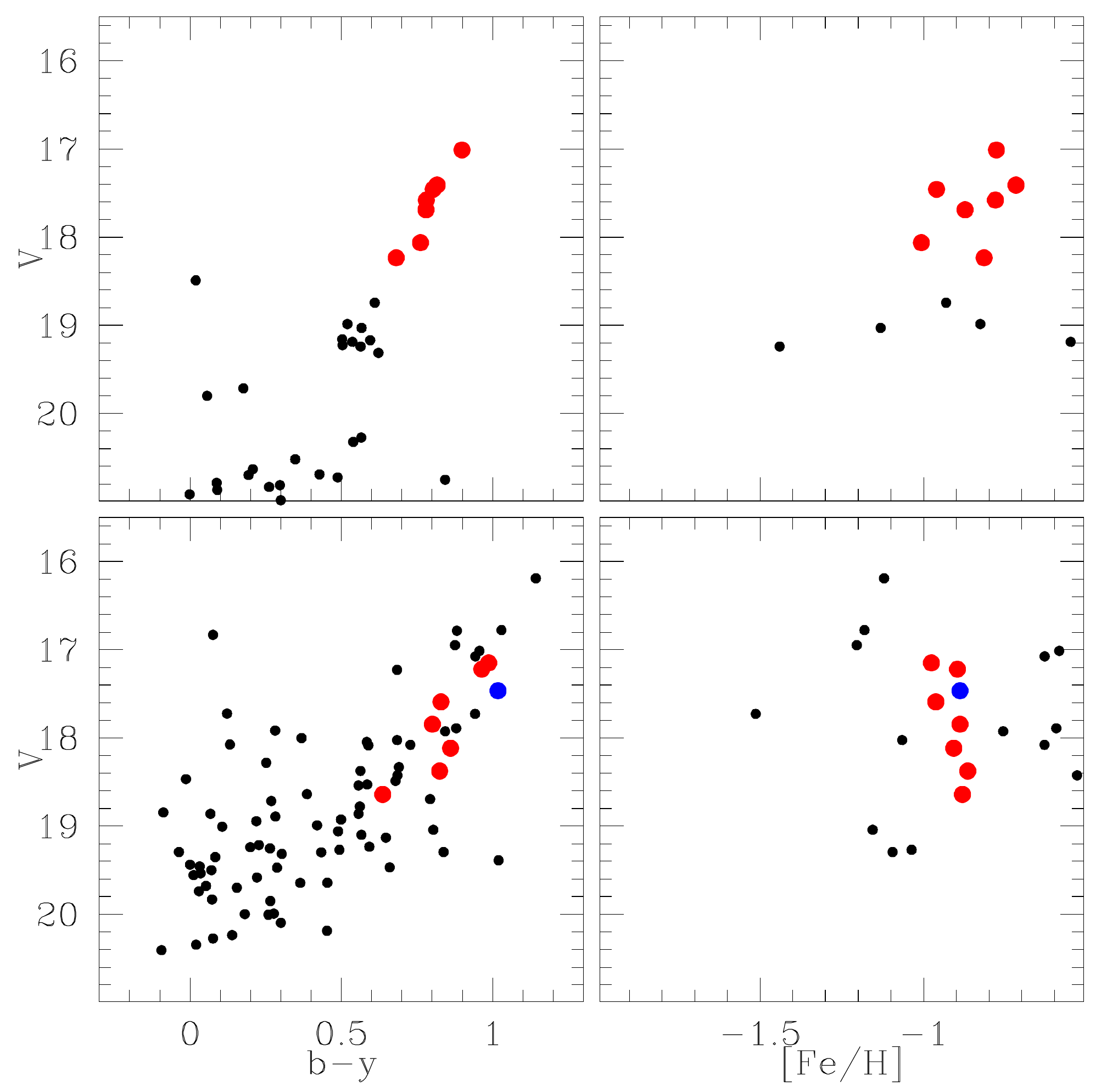}
\caption{Same as Figure~\ref{fig:fig1}, for SL\,8 (top panels) and 363 (bottom panel).}
\label{fig:fig3}
\end{figure}

\begin{figure}
\includegraphics[width=\columnwidth]{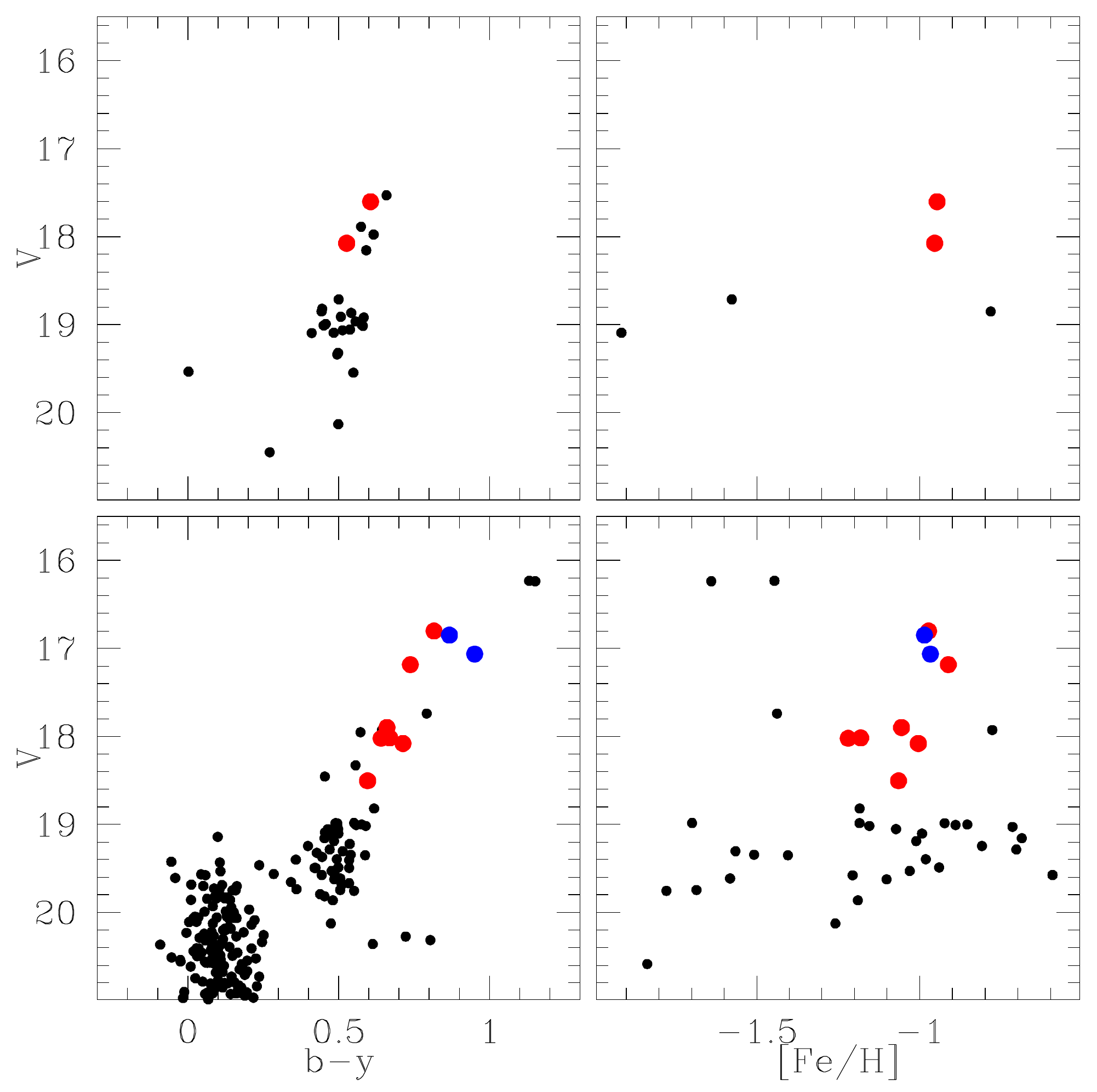}
\caption{Same as Figure~\ref{fig:fig1}, for SL\,388 (top panels) and 509 (bottom panel).}
\label{fig:fig4}
\end{figure}

\begin{figure}
\includegraphics[width=\columnwidth]{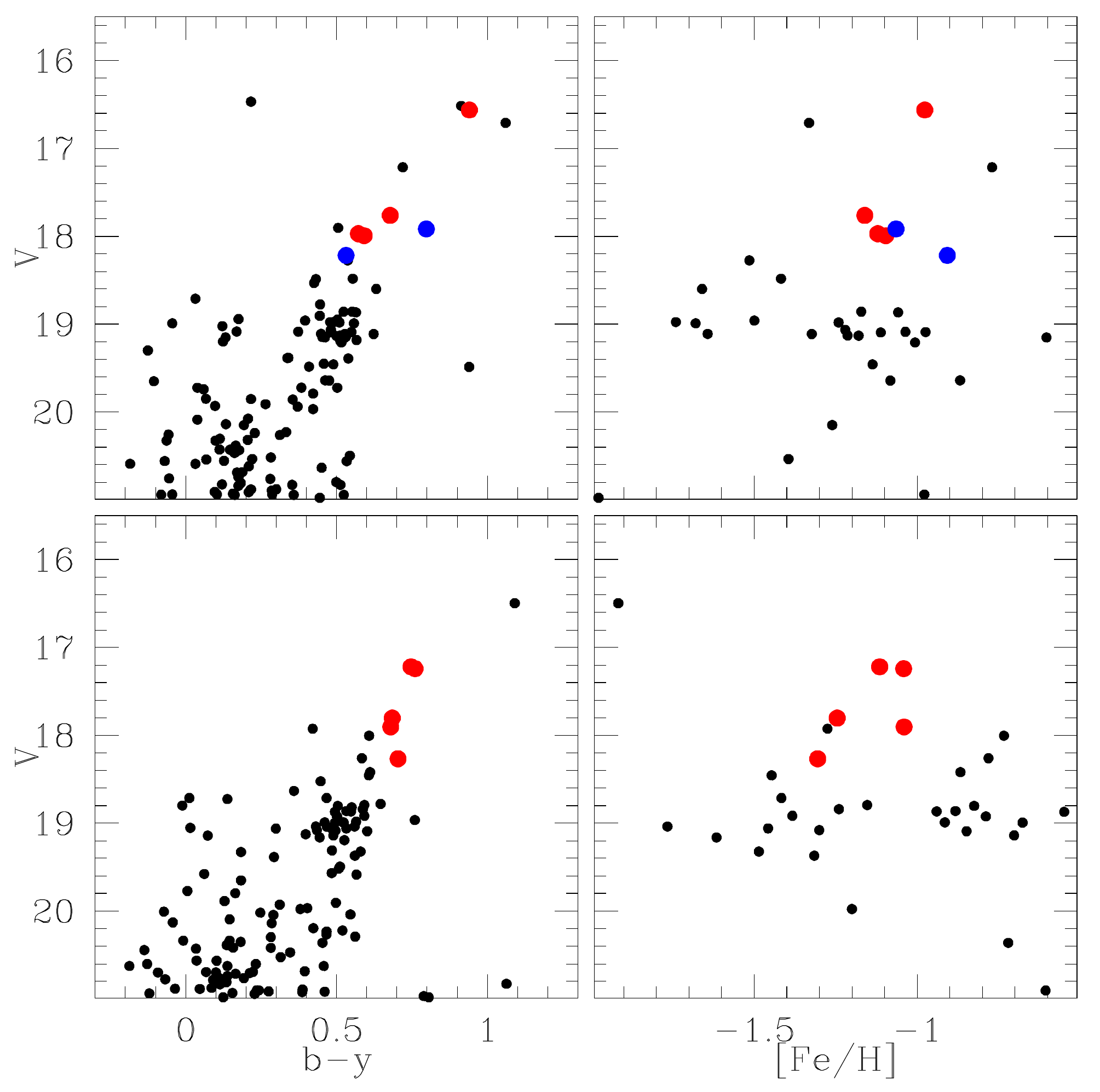}
\caption{Same as Figure~\ref{fig:fig1}, for SL\,549 (top panels) and 555 (bottom panel).}
\label{fig:fig5}
\end{figure}

\begin{figure}
\includegraphics[width=\columnwidth]{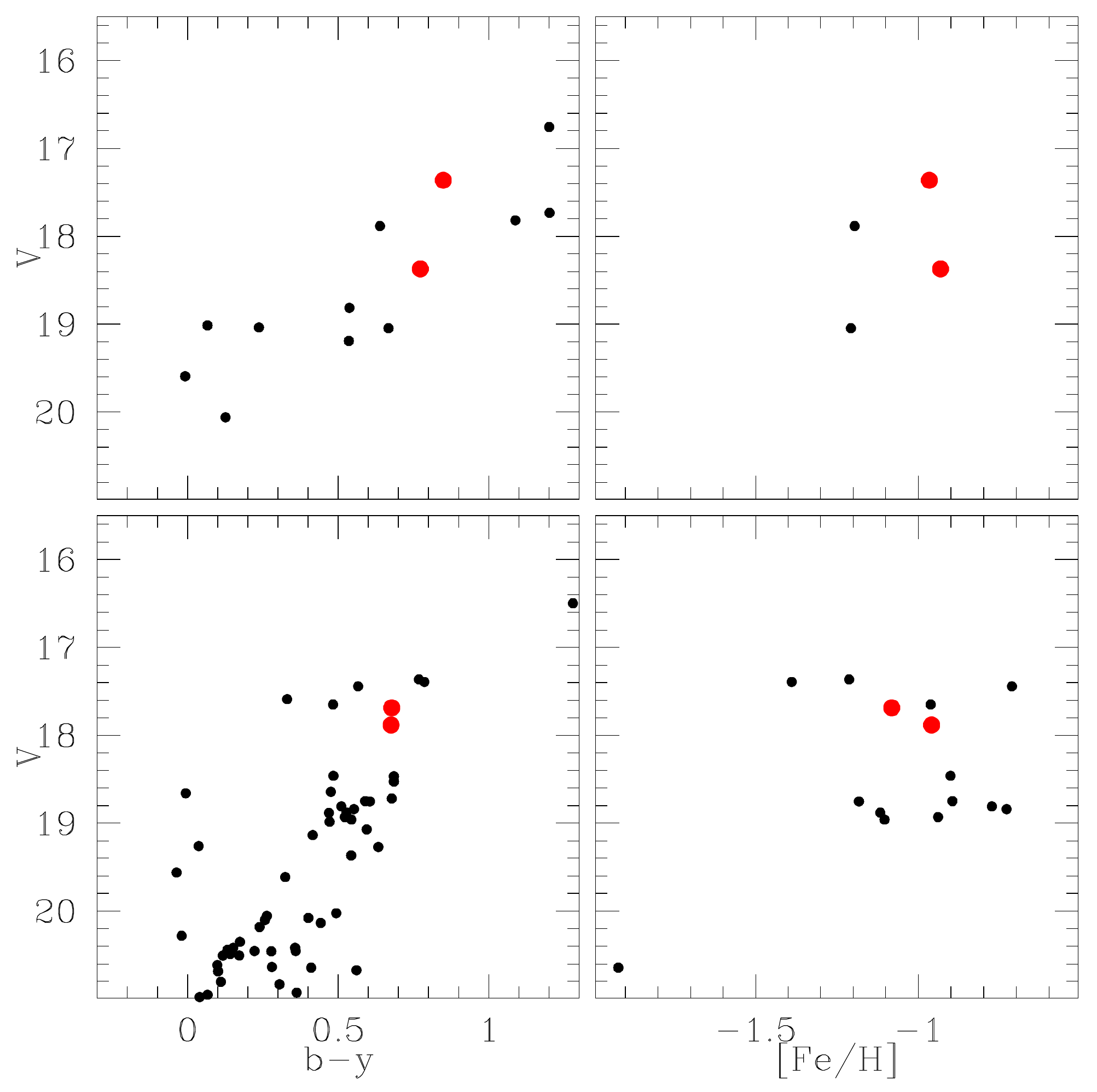}
\caption{Same as Figure~\ref{fig:fig1}, for SL\,817 (top panels) and 842 (bottom panel).}
\label{fig:fig6}
\end{figure}

\begin{figure}
\includegraphics[width=\columnwidth]{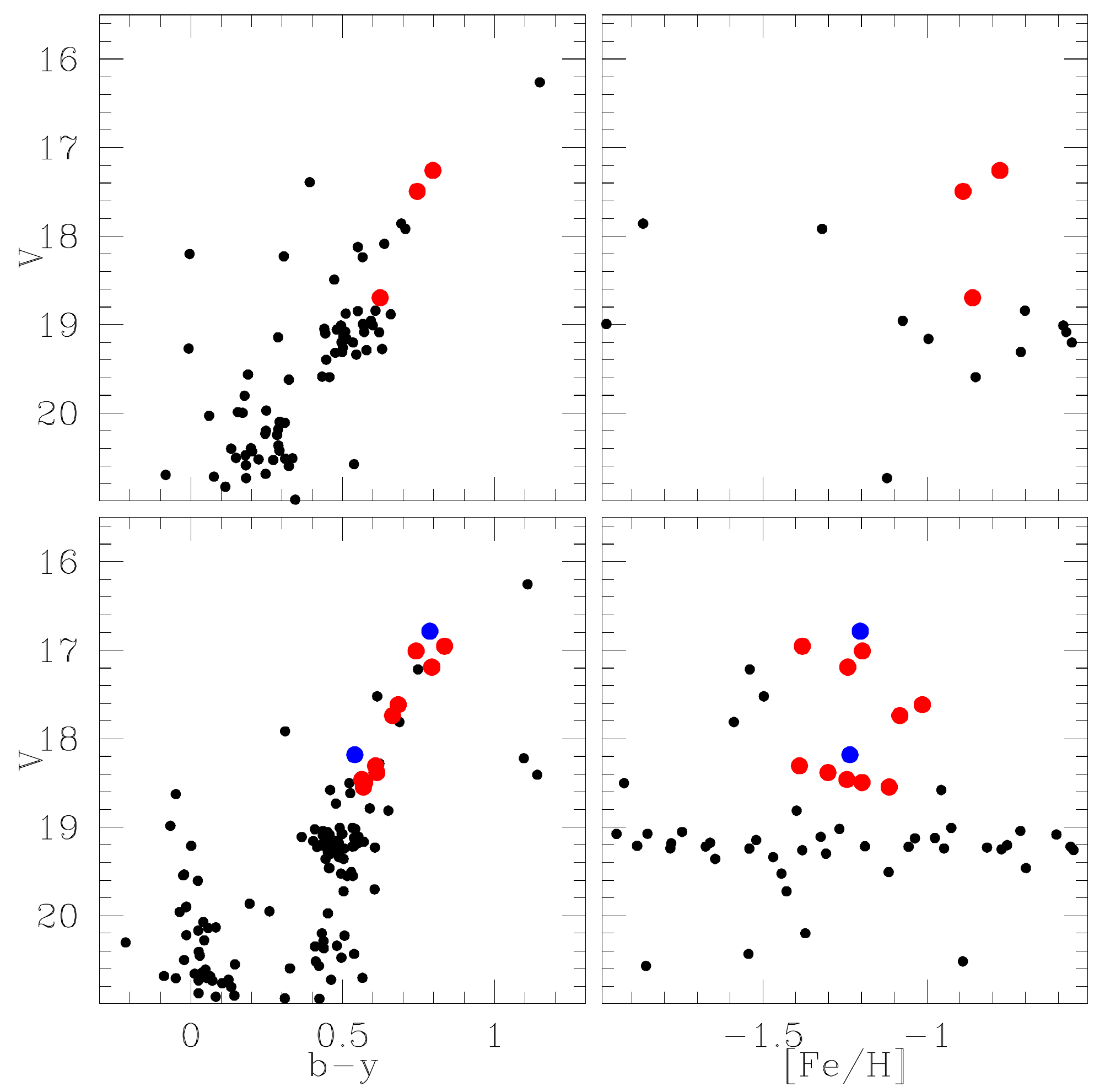}
\caption{Same as Figure~\ref{fig:fig1}, for SL\,862 (top panels) and L\,6 (bottom panel).}
\label{fig:fig7}
\end{figure}

\begin{figure}
\includegraphics[width=\columnwidth]{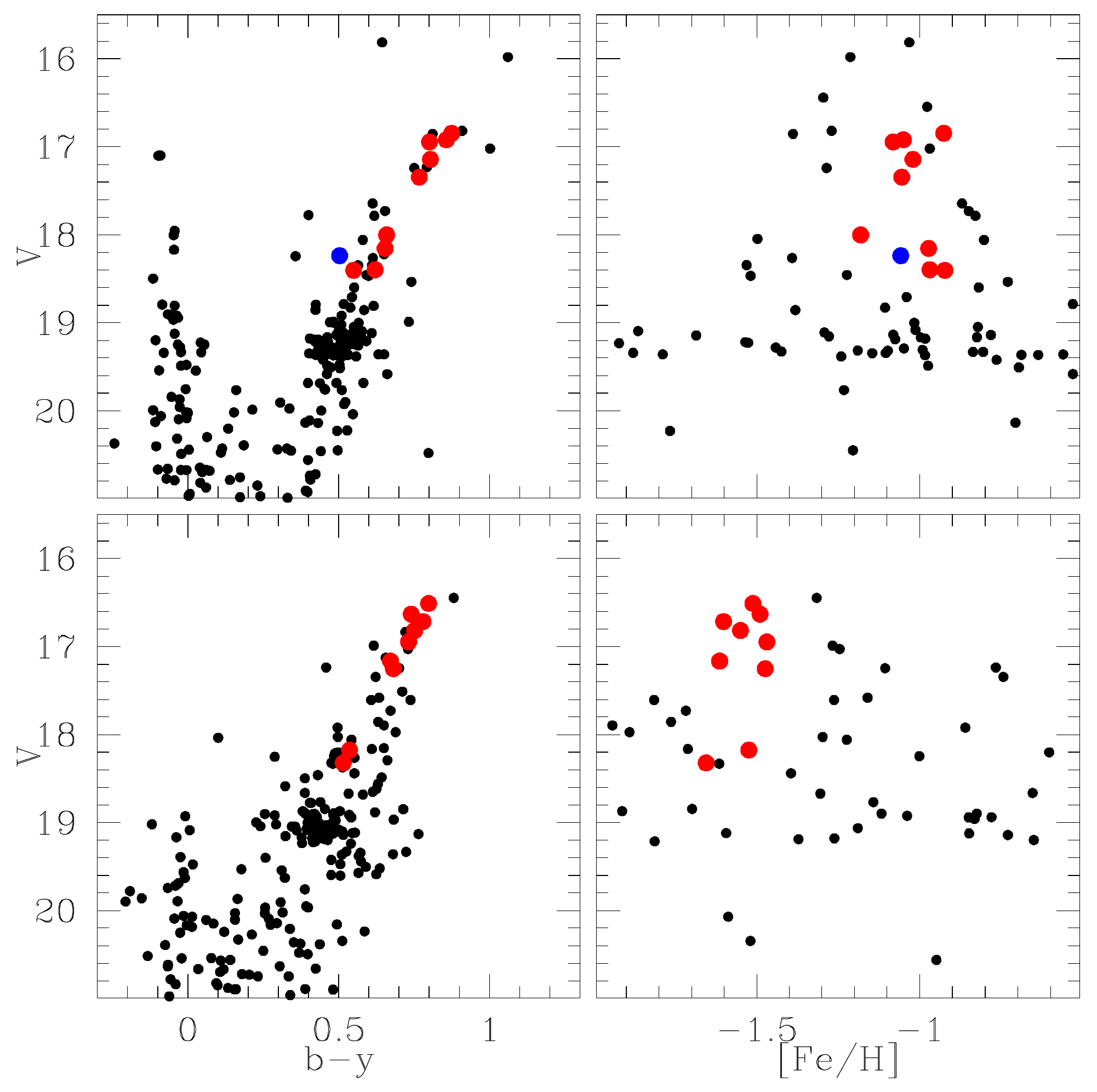}
\caption{Same as Figure~\ref{fig:fig1}, for L\,19 (top panels) and 27 (bottom panel).}
\label{fig:fig8}
\end{figure}

\section{Metallicity estimates}

The metallicity calibration obtained by \citet{calamidaetal2007} involves the $m_1$$_0$ and $(v-y)_0$ 
indices  of red giant branch stars, corrected by reddening, according to the following expression: 

\begin{equation}
{m_{\rm 1}}_o = \alpha + \beta {\rm [Fe/H]} + \gamma (v-y)_o + \delta {\rm [Fe/H]}(v-y)_o
\end{equation}

\noindent where $\alpha$ = -0.309, $\beta$=-0.090$\pm$0.002, $\gamma$=0.521$\pm$0.001, and
$\delta$=0.159$\pm$0.001, respectively. While de-reddening color indices is a somehow 
straightforward task, the selection of red giant branch stars requires several steps. We retrieved the 
$E(B-V)$ values obtained by \citet{sf11} from the NASA/IPAC Infrared Science  
Archive\footnote{https://irsa.ipac.caltech.edu/} (see Table~\ref{tab:tab1}) using the star
clusters' coordinates, and computed the intrinsic colors:\\

$(v-y)_0 = (v-y) - 1.67\times 0.74E(B-V)$\\

\noindent and\\

$m_1$$_0$$ = (v-b) - (b-y) + 0.33 \times 0.74E(B-V)$\\

\noindent where the reddening laws,  $E(X)/E(B-V)$, with $X = v-y,  m_1$, are those given by \citet{cm1976}. The $(v-y)_0$ and $m_1$$_0$ values were then entered in eq. (4) to compute
individual stellar metallicities. Their errors were estimated from a full analytical propagation of errors,
including those on the calibration coefficients, as follows:\\

\newpage

$\sigma({\rm [Fe/H]})^2 = \left(\frac{\partial {\rm [Fe/H]}}{\partial \alpha} \sigma(\alpha)\right)^2 
+ \left(\frac{\partial {\rm [Fe/H]}}{\partial \beta} \sigma(\beta)\right)^2 +
\left(\frac{\partial {\rm [Fe/H]}}{\partial \gamma}\sigma(\gamma)\right)^2 + 
  \left(\frac{\partial {\rm [Fe/H]}}{\partial \delta}\sigma(\delta)\right)^2 + 
  \left(\frac{\partial {\rm [Fe/H]}}{\partial {m_{\rm 1}}_o}\sigma({m_{\rm 1}}_o)\right)^2 + 
\left(\frac{\partial {\rm [Fe/H]}}{\partial (v-y)_o}\sigma((v-y)_o)\right)^2$,\\

\vspace{0.5cm}

$\sigma({\rm [Fe/H]})^2 = \left(\frac{0.002{\rm [Fe/H]}}{c}\right)^2 + 
\left(\frac{0.001(v-y)_o}{c}\right)^2 + \left(\frac{0.001{\rm [Fe/H]}(v-y)_o}{c}\right)^2 +
\left(\frac{\sigma({m_{\rm 1}}_o)}{c}\right)^2 + 
\left(\frac{(-0.521c - 0.159({m_{\rm 1}}_o +0.309 -0.521(v-y)_o)\sigma((v-y)_o)}{c^2}\right)^2$,

\vspace{0.5cm}

\noindent where c = $-0.090 + 0.159(v-y)_o$, and $\sigma({m_{\rm 1}}_o)$ and $\sigma((v-y)_o)$
are the photometric errors in ${m_{\rm 1}}_o$ and $(v-y)_o$, respectively. The mean star clusters'
metallicities were calculated using the individual [Fe/H] values of the selected stars, weighed by their
respective uncertainties (see   last column of  Table~\ref{tab:tab1}).  We provide in the Appendix 
A a comparison of 
the [Fe/H] values derived for the selected stars (red filled circles in Figures~\ref{fig:fig1}-\ref{fig:fig8}) from 
the semi-empirical calibration with those based on the empirical and theoretical ones, respectively.

Star clusters' red giant branch stars were selected according to the following criteria: (i) The stars are
located inside the clusters' radii \citep{hz2006,betal08,wz11}. This is a basic starting point, because
contamination from field stars is also present inside those areas. Furthermore, intermediate-age and
old LMC/SMC star clusters are frequently projected toward star fields with similar ages and metallicities 
\citep{getal03,p11c}, which means that they also populate the star cluster CMD regions. (ii)  They
are distributed along the red giant branch and above the star cluster red clump/horizontal branch in the
$V$ versus $b-y$ CMD; $b-y$ is mainly a temperature effective indicator with less metallicity 
sensitivity \citep{cm1976}. We initially imposed the simple restriction: $V <$ 18.7 mag and $b-y >$ 0.4
mag (see Figures~\ref{fig:fig1}-\ref{fig:fig8}, left panels). These magnitude and color cuts still allow 
several field stars appear not only along the star clusters'  red giant branches, but also far away of 
them. Stars outside the red giant branches were later easily discarded when applying the metallicity
cut. (iii) The star clusters' red giant branch stars span a readily visible range of [Fe/H] values in the $V$
versus [Fe/H] plane. Figures~\ref{fig:fig1}-\ref{fig:fig8} (right panels) highlight them with red large filled 
circles, where all the stars distributed within the star  clusters' radii that have a metallicity estimate from
eq. (4) are also shown with black dots, irrespective if they are red giant branch stars  (see also 
Appendix C).

\begin{figure}
\includegraphics[width=\columnwidth]{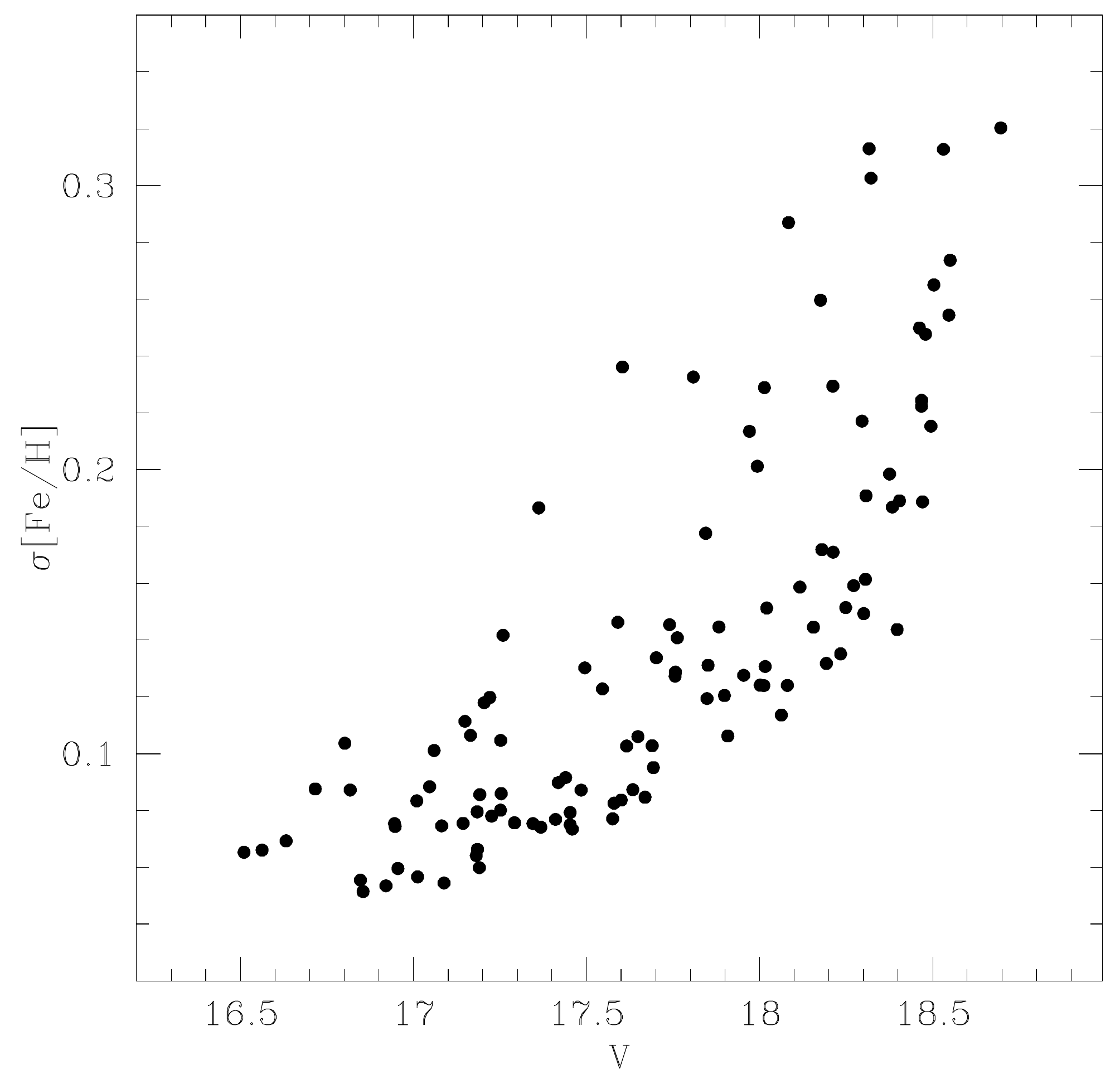}
\caption{Individual metallicity errors as a function of the $V$ magnitude of stars selected in star
cluster, represented by red large filled circles in Figures~\ref{fig:fig1}-\ref{fig:fig8}.}
\label{fig:fig9}
\end{figure}

While identifying the star clusters' red giant branch candidates we considered the [Fe/H] errors 
($\sigma$[Fe/H]). Figure~\ref{fig:fig9} shows $\sigma$[Fe/H] as a function of $V$ for all the selected
stars. As can be seen, the fainter a star the larger $\sigma$[Fe/H]. For this reason, we gave more 
weight to the brightest stars with more accurate photometry and hence with the smallest [Fe/H] 
uncertainties.  Likewise, as the $V$ magnitude increases, both  $\sigma$[Fe/H] and the dispersion of 
the individual [Fe/H] values increase, because of the poorer photometry quality, so that the observed
[Fe/H] range of the brightest selected stars more properly reveals the star cluster metallicity range. 
This metallicity range should be the same at any magnitude level. We also beard in mind that stars 
distributed along or adjacent to the star clusters' red giant branches could be field stars with ages
difference from those of the star clusters (and hence difference metallicities). These stars 
spuriously produce a wider spread of [Fe/H] values for relatively bright $V$ magnitudes. The combined
population of field and star cluster red clump stars also produce wide ranges of [Fe/H] values. 
Here we did not use those stars and kept them in Figures~\ref{fig:fig1}-\ref{fig:fig8} for illustrative
purposes.

\section{Analysis}

\begin{figure}
\includegraphics[width=\columnwidth]{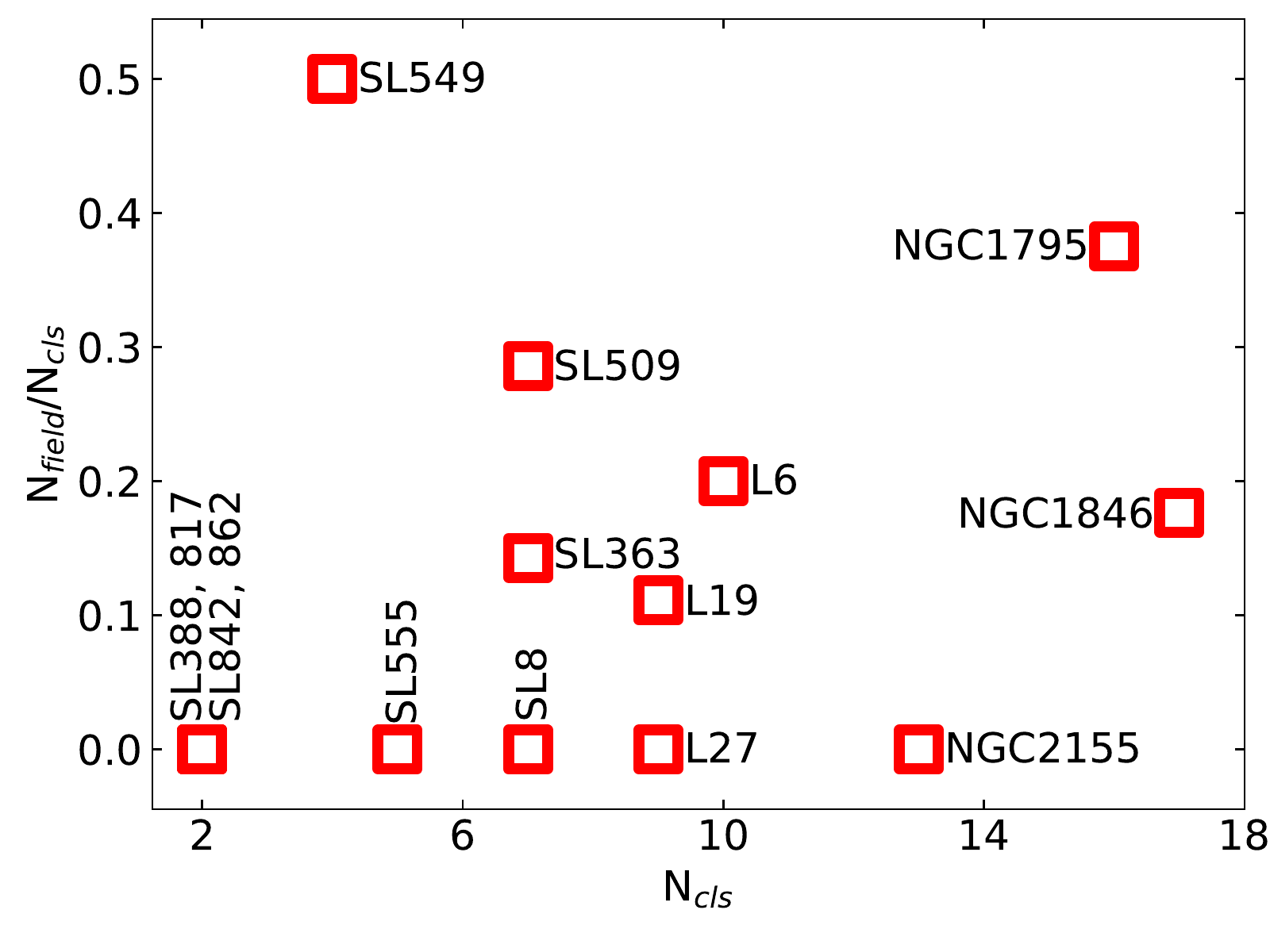}
\caption{Ratio of the number of stars in the field and that in the star clusters versus the number of
selected stars in each star cluster.}
\label{fig:fig10}
\end{figure}

We evaluated the degree of contamination by field stars in the resulting mean star cluster metallicities.
We are interested in those stars that lie along the star clusters' red giant branches and also have 
metallicities close to the star clusters' values. These stars play a role when computing the mean
star cluster metallicities, because we cannot distinguish them from star clusters' stars without
the availability of proper motions or radial velocities. We used different equal star clusters areas
located reasonably far from the star clusters and counted the number of stars 
(N$_{field}$, represented with blue filled circles in Figures~\ref{fig:fig1}-\ref{fig:fig8}) distributed
inside them that satisfy the criteria required for the selected red giant branch stars (N$_{cls}$). We then
computed the ratio N$_{field}$/N$_{cls}$ and built Figure~\ref{fig:fig10}, which depicts its variation as a
function of N$_{cls}$. As can be seen, half of the star cluster sample is affected by a mean null field
star contamination. SL\,549 amounts 50$\%$ of contribution from field stars in the derived mean
metallicity. Nevertheless, Figure~\ref{fig:fig5} (right panel) shows that this is because of the small
number of stars used. These field stars could increase the standard deviation up to $\sim$ 0.03 dex,
the mean metallicity value would not change. A similar interpretation could be applied to the remaining
stars  clusters, where the smaller the N$_{field}$/N$_{cls}$ ratio, the smaller the increase of the respective standard deviation. 

\begin{figure}
\includegraphics[width=\columnwidth]{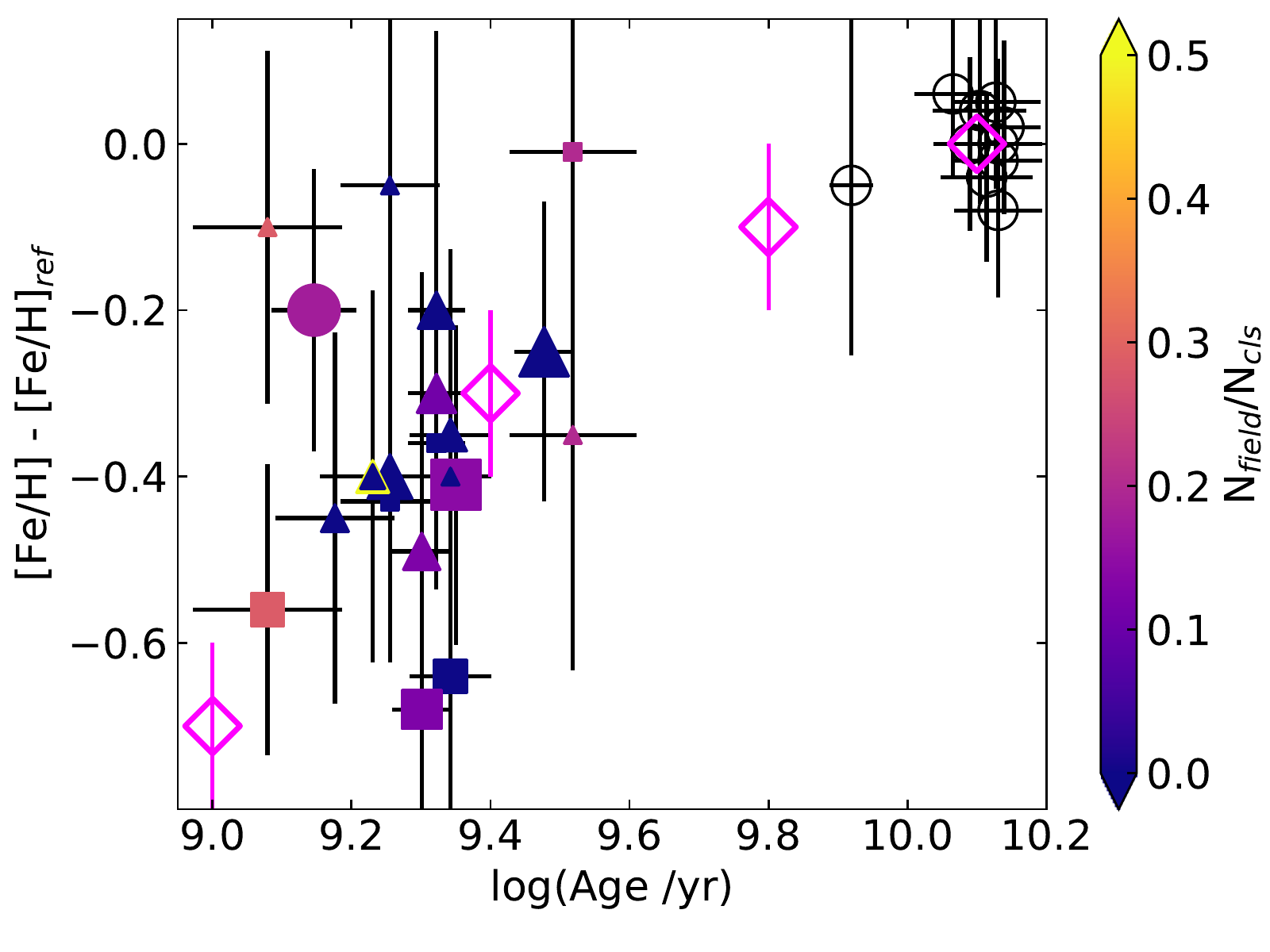}
\caption{Metallicity difference as a function of the star cluster's age. Filled circles, squares, and
triangles correspond to high-resolution, low-resolution spectroscopy, and photometry reference values,
respectively. Open magenta diamonds represent metallicity differences derived from theoretical
isochrones (see text for details).}
\label{fig:fig11}
\end{figure}

\begin{figure}
\includegraphics[width=\columnwidth]{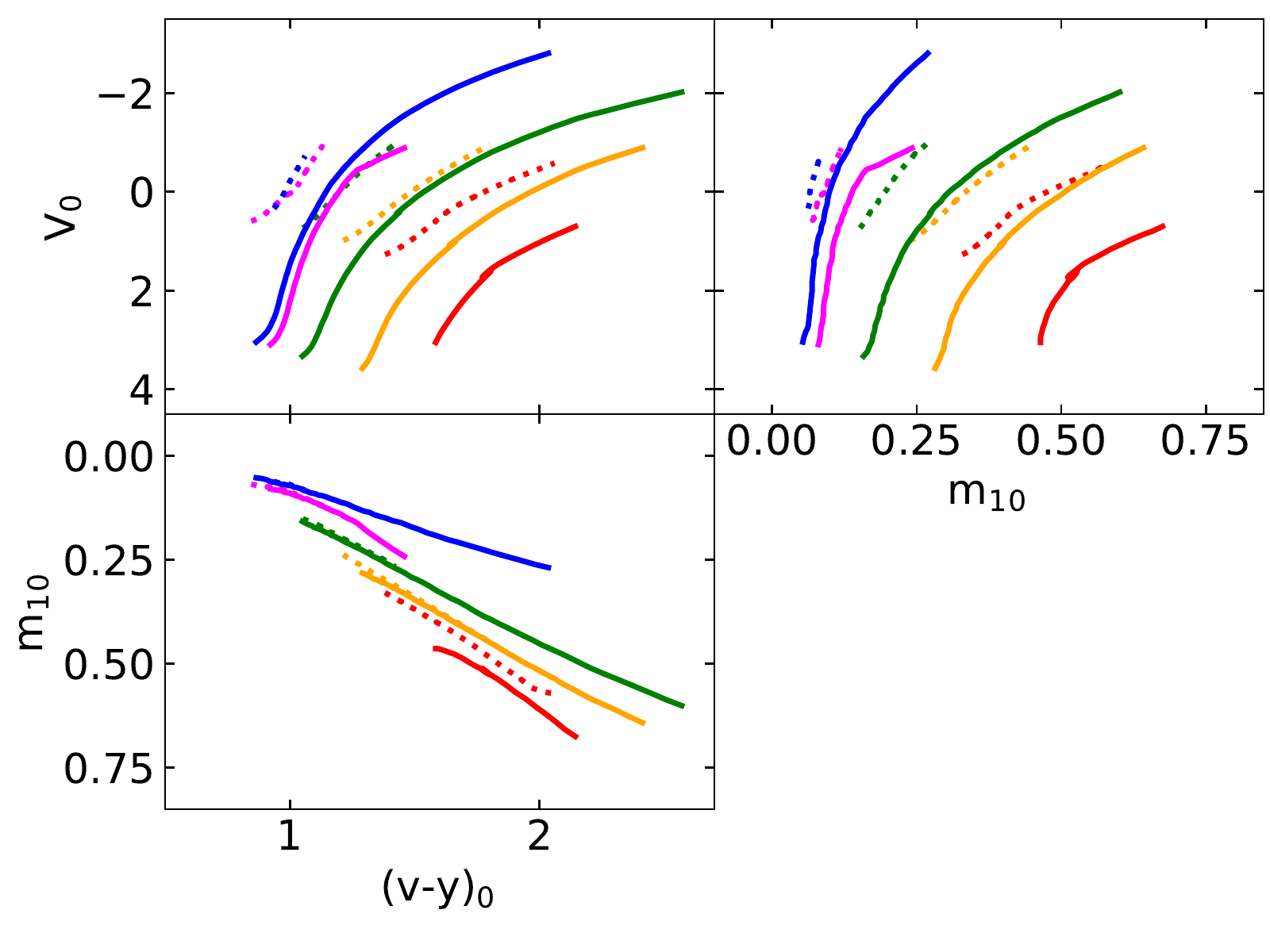}
\caption{Theoretical red giant branches for log(age /yr)=9.0 (dotted lines) and 10.10 (solid lines) and
[Fe/H] = -2.0, -1.5, -1.0, -0.5, and 0.0 dex shown with blue, magenta, green, orange, and red colors,
 respectively.}
\label{fig:fig12}
\end{figure}

\begin{figure}
\includegraphics[width=\columnwidth]{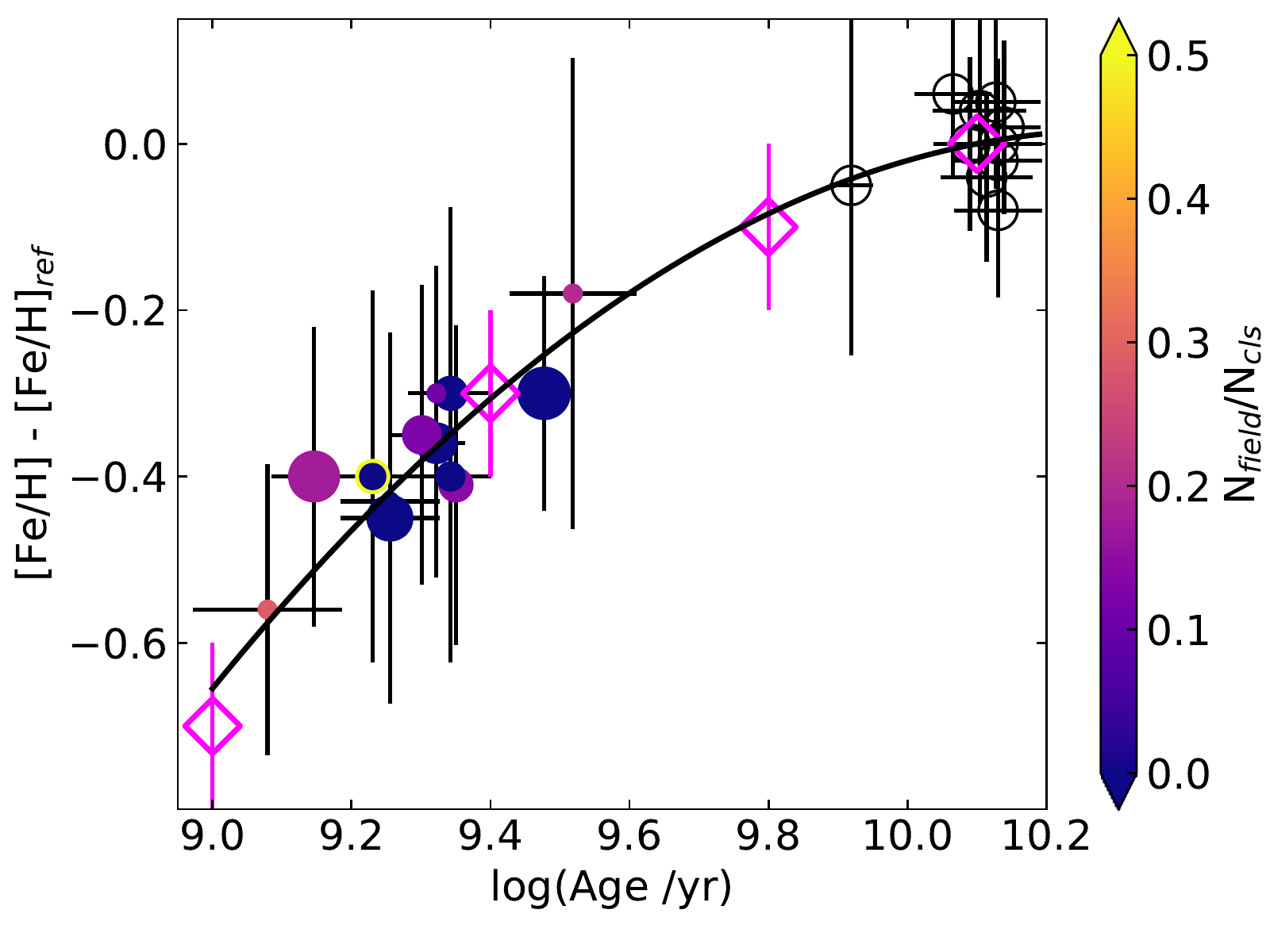}
\caption{Same as Figure~\ref{fig:fig11} for adopted weighted mean [Fe/H] values.
The solid line represents the least squared fit obtained (see text for details).}
\label{fig:fig13}
\end{figure}

We searched the literature for ages and metallicities of the studied star cluster sample. 
Appendix B deals with the individual values available in the literature, while the adopted 
values are listed in Table~\ref{tab:tab1}. From that piece of information  
 (see Table~\ref{tab:tab1appendix})
we computed the difference between the  individual reference star clusters' metallicities and the 
values derived in this work,  distinguishing those coming from high- or low-resolution spectroscopy, or photometry. Figure~\ref{fig:fig11} shows the relationship between the resulting metallicity differences
 (present - reference) and the star clusters' ages.  We restricted to reference [Fe/H] values lower 
than -0.5 dex, included their associated errors, so that to comply with the metallicity range of eq. (4). We used all clusters listed in Table~\ref{tab:tab1}, except  NGC\,1795 and SL\,817.
The point sizes are proportional to N$_{cls}$, while their error bars come from adding in quadrature the uncertainties of the  reference (Table~\ref{tab:tab1appendix}) and present mean
[Fe/H] values included in Table~\ref{tab:tab1}. The points are colored according to the
N$_{field}$/N$_{cls}$ ratios. Figure~\ref{fig:fig11} reveals that the Str\"omgren metallicities derived from
eq.(4) are age-dependent, and therefore, they need to be corrected for further use.  A general
trend is readily visible, which points of the need of discarding some individual discrepant [Fe/H] values.
In order to have a second reference source, we used the [Fe/H] values derived from Str\"omgren
theoretical isochrones \citep[][$Y$ = 0.2485 + 1.78$Z$, $Z_\odot$= 0.0152]{betal12} to superimpose them onto Figure~\ref{fig:fig11} (open magenta 
diamonds). We first built $V$ vs. $(v-y)_0$ and $V$ vs. $m_1$$_{0}$ CMDs for [Fe/H] = -2.0,
 -1.5, -1.0, -0.5, and 0.0 dex and  log(age /yr)=10.10,  which we called standard giant branches. 
We only drew the red giant branch track sections of these five theoretical isochrones. Then, we superimposed
onto these standard giant branches the theoretical red giant branches of isochrones with  log(age /yr) = 9.0
and the five [Fe/H] values as for  log(age /yr)=10.10 (see Figure~\ref{fig:fig12}). Because of the 
age-metallicity degeneracy, the 
log(age /yr)=9.0  red giant branches appear shifted toward bluer colors with respect to the corresponding
standard red giant branches according to their metallicities. We estimated by interpolating in
the five standard giant branches, the [Fe/H] values of the  log(age /yr)=9.0 red giant branches, and
computed the difference between the obtained interpolated [Fe/H] values and those given by the respective
isochrones ([Fe/H] = -2.0, -1.5, -1.0, -0.5, and 0.0 dex).  We repeated this procedure for 
log(age /yr)=9.4 and 9.8. By using these theoretical points and the general trend observed in 
Figure~\ref{fig:fig11}, we adopted weighted mean [Fe/H] values for each cluster (see Table~\ref{tab:tab1}). 
Figure~\ref{fig:fig13} depicts the resulting distribution using the adopted mean metallicities.

With the aim of providing with  an expression for the metallicity correction that covers a wide age
range, we added ten LMC old globular clusters studied by \citet{pk2018} from Str\"omgren photometry.
We took from that study the star clusters' ages and both reference and Str\"omgren metallicities, with
their respective uncertainties. They are represented in  Figures~\ref{fig:fig11} and \ref{fig:fig13} with open circles. When performing a least square fit to all the data points, we obtained the following expression:\\

$\Delta$[Fe/H] (dex) = -41.80($\pm$43.48) + 8.11($\pm$9.00)$\times$log(age /yr)  
- 0.39($\pm$0.46)$\times$log(age /yr)$^2$,\\

\noindent with a standard deviation, a correlation factor, and an F-test coefficient of 0.05, 0.98, and
0.97, respectively. Str\"omgren metallicities calculated from eq.(4) should be corrected by an amount 
equal to $-\Delta$[Fe/H].

\section{Summary}

Based on our previous knowledge that metallicities derived from standard giant branches in
several photometric systems are affected by the age-metallicity degeneracy for star clusters younger than the old globular clusters, and bearing in mind the fundamental role that metallicities play in
our understanding of the formation and chemical evolution processes at different Universe's scales,
we decided to investigate it for the Str\"omgren $vby$ bandpasses. Until now, it has been
straightforward to compute [Fe/H] values from the Str\"omgren reddening free $(v-y)_0$ and 
$m_1$$_0$ indices; the latter being a metallicity sensitive index. In general, they have been used for 
the study of globular cluster metallicity distributions \citep[e.g.,][]{calamidaetal2014,franketal2015}, 
to unveil multiple stellar populations \citep[e.g.,][]{carrettaetal2011,massarietal2016} rather than to 
estimate the metal content of intermediate-age star clusters.

In order to probe whether such an age dependence also affects the Str\"omgren metallicities, we
made use of unexploited publicly available Str\"omgren $vby$ images of intermediate-age and old 
LMC/SMC star clusters, which have been previously targeted for independent metallicity 
analysis. Once we properly proceeded the observational material and standardized the resulting
instrumental photometry, we computed individual stellar [Fe/H] values using the semi-empirical
\citet{calamidaetal2007}'s calibration. We performed a sound selection of star cluster red giant branch
candidates in order to comply with the metallicity calibration validity. In doing this, we constrained
the cluster stars to be distributed inside the star clusters' radii, located along the star cluster red giant
branch, and with [Fe/H] values  within the readily visible star cluster metallicity range. We further
assessed the effect in the derived mean star cluster metallicities of the contamination of field stars
that are indistinguishable without proper motions or radial velocity measurements from the selected
stars used to compute those mean values.

The resulting mean [Fe/H] values and the estimated uncertainties were then compared with those
metallicities taken from the literature. We found that the measured Str\"omgren metallicities need to be
corrected by an amount that depends on the star cluster age, in the sense that, the younger the star 
cluster, the larger the metallicity correction. Since the measured [Fe/H] values result more metal-poor
than the reference [Fe/H] ones, a positive correction should be added to the former one. From 26
relative metallicity points spanned from $\sim$ 1 Gyr up to the globular cluster ages, we fitted
a quadratically age-varying curve that provides with metallicity corrections with an overall uncertainty
of $\sim$ 0.05 dex. We finally repeated a similar comparison from a fully independent approach, which consisted in using theoretical red giant branches to trace the standard red giant branches and those of 
star clusters with ages in the aforementioned  age range. Our findings show a very good agreement 
between the measured relative metallicities and those derived from theoretical isochrones.

\begin{acknowledgements}
I thank the referee for the thorough reading of the manuscript and
timely suggestions to improve it. 
%A.E.P. acknowledge support from the Ministerio de Ciencia, Tecnolog\'{\i}a e Innovaci\'on Productiva (MINCyT) through grant PICT-201-0030. 
\end{acknowledgements}

%\bibliographystyle{aa}
%\bibliography{paper} % if your bibtex file is called paper.bib

%\input{paper.bbl}

\begin{appendix} 

\section{[Fe/H] values based on \citet{calamidaetal2007}'s calibrations}

\begin{figure}
\includegraphics[width=\columnwidth]{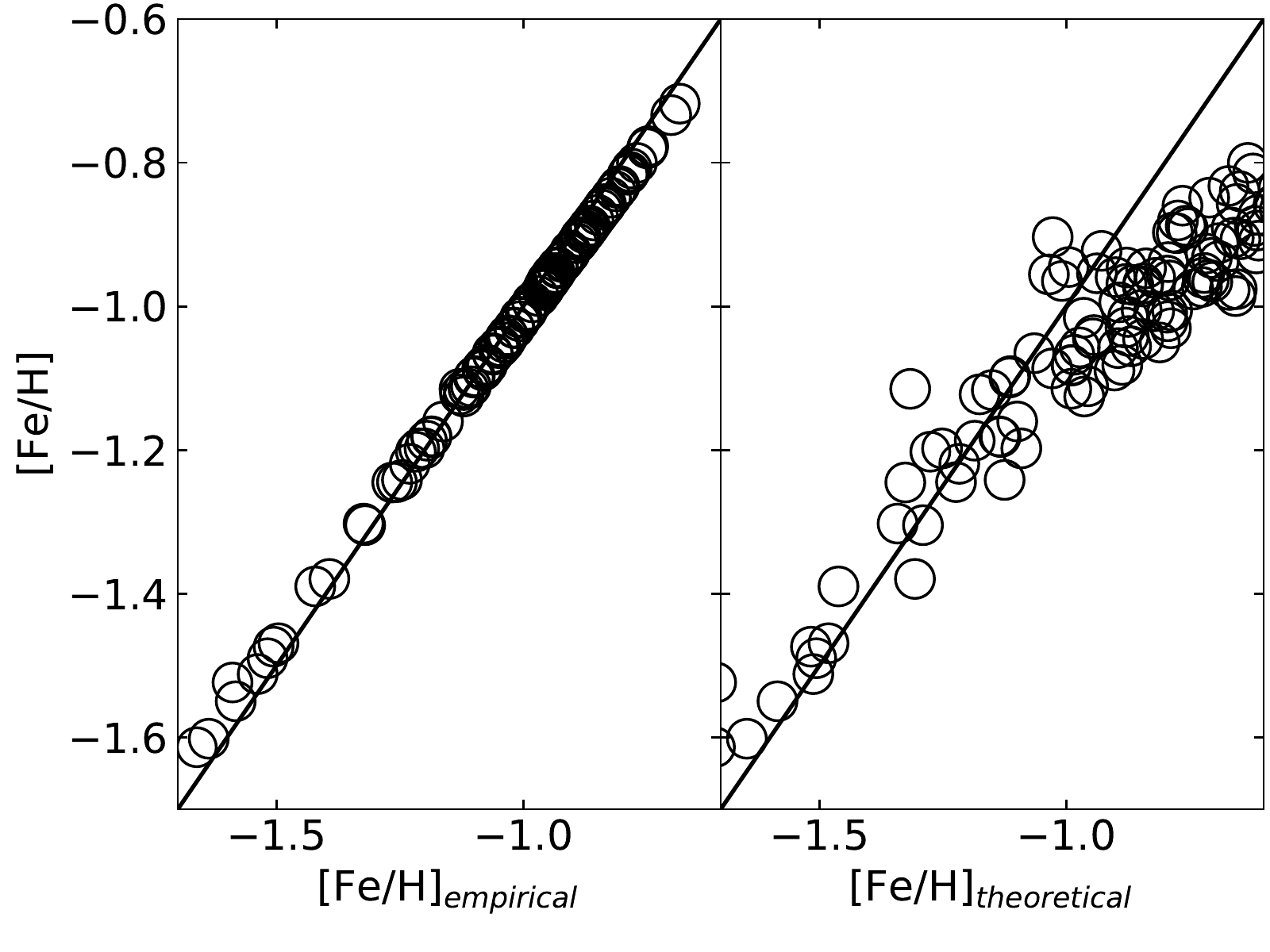}
\caption{[Fe/H] values from eq. 4 compared with those obtained from the empirical (left panel) and 
the theoretical (right pane) calibrations of \citet{calamidaetal2007}.The solid line represents the identity relationship.}
\label{fig:fig1appendix}
\end{figure}

For the sake of the reader, we present here a comparison between metallicities derived from different 
\citet{calamidaetal2007}'s calibrations, namely: empirical, theoretical, and semi- empirical (eq. 4) ones. 
As can be seen in Figure~\ref{fig:fig1appendix}, metallicities derived with the empirical or semi-empirical calibrations agree quite well. 
However, metallicities derived with the theoretical calibration are systematically more metal-poor compared to metallicities derived with the empirical and semi-empirical calibrations for [Fe/H] $>$ -1.0 dex.

\section{Cluster metallicities}

We searched the literature looking for metallicity estimates for the cluster sample. Because of the
relatively low brightness and small angular size of some clusters, photometric estimates were
more frequently found,  although some low-resolution spectroscopy values and other few 
high-spectroscopy ones were found. When comparing between them we found in some cases a 
significant dispersion,  independently of the observational technique used (details are at the bottom of 
Table~\ref{tab:tab1appendix}), which prevented us to favor any particular technique. In order
to adopt weighted mean values, we considered the whole collection of [Fe/H] values (see
Figure~\ref{fig:fig11}), which allowed us to recognize a general trend, and therefore to  discard 
some visibly discrepant metallicities. We would like to note that possible sources of uncertainty due to different studies using different metallicities scales might exist.

\begin{table*}
\caption{[Fe/H] values for the cluster sample.}
\label{tab:tab1appendix}
\begin{tabular}{@{}lcclcclcc}\hline\hline
Star cluster & [Fe/H] & Reference & Star cluster & [Fe/H] & Reference & Star cluster & [Fe/H] & Reference \\\hline
NGC\,1651 & -0.37$\pm$0.20 & 1 & SL\,8		& -0.40$\pm$0.20 & 14 & SL\,862	&-0.47$\pm$0.07	&15\\
		& -0.82$\pm$0.44 & 2 & 		& -0.50$\pm$0.30 & 20 & 		&-0.85$\pm$0.20 & 14\\
		& -0.07$\pm$0.10 & 3 & 		&		&     & 		&		&     \\
		& -0.63 to -0.45 & 4 & SL\,363	& -0.49$\pm$0.12& 15 & L\,6	&	-1.24$\pm$0.03	&17\\
		& -0.30$\pm$0.03 & 5 & 		&		&     & 		&-0.90$\pm$0.20 & 18 \\
		&-0.70$\pm$0.10 & 6 & SL\,388	& -0.65$\pm$0.20	&14& 		&		&     \\
		&		&     & 		& -0.58$\pm$0.06 & 21 & L\,19	&	-0.75$\pm$0.10&18\\
NGC\,1795 & -0.40$\pm$0.10 & 12	& 		&		&     & 		&		&     \\
                   & -0.23$\pm$0.20 &1 & SL\,509	&-0.54$\pm$0.09	&15& L\,27	&	-1.14$\pm$0.06&17\\
		&		&     & 		&-1.18$\pm$0.09 & 21 & 		&-1.30$\pm$0.30 & 18 \\
NGC\,1846 &	-0.50$\pm$0.10 & 13 & 		&-0.85$\pm$0.30 & 20 &  &	&  \\
		& -0.76$\pm$0.20 & 1 & 		&		&     &&	&  \\
		& -0.70$\pm$0.08 & 19 & SL\,549	&-0.70$\pm$0.20	&9 &&	&  \\
		&		&      & 		&		&     &&	&  \\
NGC\,2155 & -0.55$\pm$0.20 & 1	& SL\,555	&-0.70$\pm$0.20	&16&&	&  \\
		& -0.60$\pm$0.20 & 7 & 		&		&     &&	&  \\
		&-0.98 to -0.80 &	 8 & SL\,817	&-0.41$\pm$0.05	&15&&	&  \\
		&-0.44$\pm$0.86	&2	&                 & -0.50$\pm$0.20 & 14&&	&  \\
		&$\sim$-0.80	& 9	& 		&		&     &&	&  \\
		&$\sim$-0.70	& 10	& SL\,842	&-0.60$\pm$0.20	&14&&	&  \\
		&$\sim$-1.0		& 11	&			&-0.36$\pm$0.20 & 1 &&	&  \\
		&-0.70$\pm$0.10 & 6 &&	&  \\\hline 
		
\end{tabular}

Reference (technique) : (1) \citet{oetal91} (low-resolution spectroscopy); (2) \citet{lr2003} (integrated 
spectroscopy); 
(3) \citet{sarajedinietal2002} (NIR photometry); (4) \citet{dirschetal2000} (Str\"omgren photometry);
(5) \citet{mucciarellietal2008} (high-resolution spectroscopy); (6) \citet{ketal07} (HST photometry);
(7) \citet{richetal2001} (HST photometry); (8) \citet{bertellietal2003} (VLT photometry); 
(9) \citet{petal03b} (Washington photometry; (10) \citet{wooetal2003} (VLT photometry); 
(11) \citet{s98} (HST photometry); (12)  \citet{getal10} (BVI photometry); (13) \citet{getal06}
(low-resolution spectroscopy); (14) \citet{betal98} (Washington photometry); (15) \citet{shetal10}
(low-resolution spectroscopy); (16) \citet{getal03} (Washington photometry); (17) \citet{parisietal2009}
(low-resolution spectroscopy); (18) \citet{petal05b} (Washington photometry); (19) \citet{songetal2019}
(high-resolution spectroscopy); (20) \citet{palmaetal2016} (Washington photometry); (21)
\citet{pieresetal2016} (DES photometry).

\end{table*}

\section{Color-color diagram of the studied clusters}

\begin{figure}
\includegraphics[width=\columnwidth]{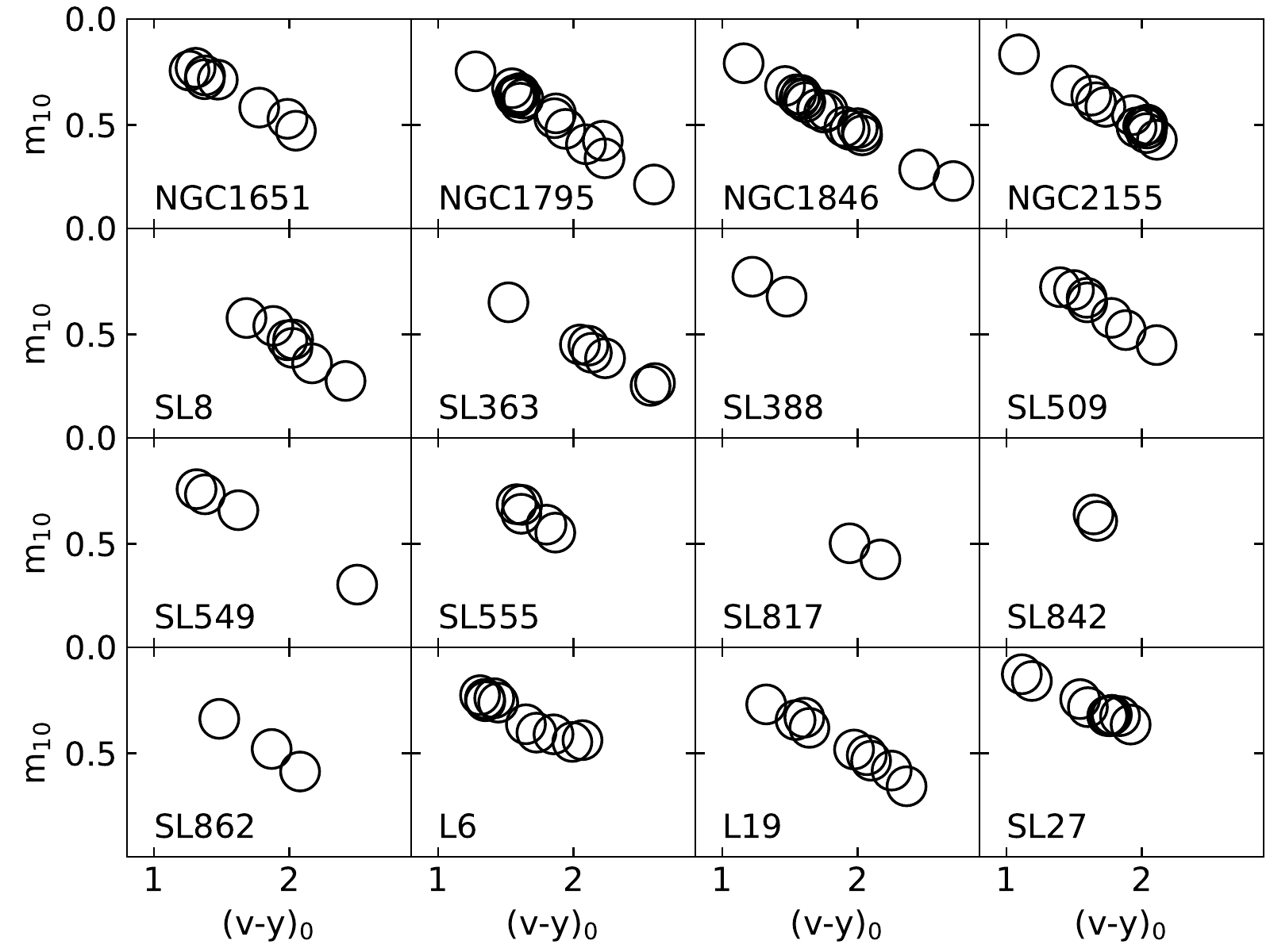}
\caption{Color-color diagram of the selected stars (red filled circles in Figures~\ref{fig:fig1}-\ref{fig:fig8}).}
\label{fig:fig2appendix}
\end{figure}

For the sake of the reader, Figure~\ref{fig:fig2appendix} shows the $m_1$$_0$ vs. $(v-y)_0$ diagrams
with the positions of the selected stars in the studied LMC and SMC clusters.

\end{appendix}

\end{document}